\documentclass[aps,pra,superscriptaddress,floatfix,nofootinbib]{revtex4}
\usepackage{mathrsfs}
\usepackage{epsfig,psfrag}

\usepackage{amsmath,amsfonts,amssymb, latexsym}
\usepackage[dvipsnames,usenames]{color}
\usepackage{subfigure}
\usepackage{bm}
\usepackage{ulem}
\newcommand{\cZ}{\mathcal Z}

\newcommand{\cC}{\mathcal C}

\newcommand{\cE}{\mathcal E}
\newcommand{\cO}{\mathcal O}
\newcommand{\cT}{\mathcal T}

\newcommand{\bx}{\mathbf x}

\newcommand{\bE}{\mathbf E}

\definecolor{BrickRed}{cmyk}{0,0.89,0.94,0.28}
\definecolor{MidnightBlue}{cmyk}{0.98,0.13,0,0.43}
\definecolor{DarkGreen}{rgb}{0,0.7,0.1}
\newcommand{\add}[1]{{\color{BrickRed} #1}}

\definecolor{RedViolet}{cmyk}{0.07,0.90,0,0.34}
\definecolor{SeaGreen}{cmyk}{0.69,0,0.50,0}
\definecolor{FireOrange}{rgb}{1.,.294,.247}

\definecolor{te}{rgb}{0.92,0.,0.65}

\newcommand{\sjrdel}[1]{{\color{ForestGreen}\sout{#1}}}

\begin{document}
\preprint{draft}
\title{Casimir potential of a compact object enclosed by a spherical cavity}%
\author{Saad Zaheer\footnote{Present address: Department of Physics and Astronomy, University of Pennsylvania, Philadelphia, PA 19104-6396, USA}}
\affiliation{Department of Physics, Massachusetts Institute of Technology, Cambridge, MA 02139, USA}
\author{Sahand Jamal Rahi}
\affiliation{Department of Physics, Massachusetts Institute of Technology, Cambridge, MA 02139, USA}
\author{Thorsten Emig}
\affiliation{Institut f\"ur Theoretische Physik, Universit\"at zu K\"oln, Z\"ulpicher Strasse 77, 50937 K\"oln, Germany}
\affiliation{Laboratoire de Physique Th\'eorique et Mod\`eles Statistiques, CNRS UMR 8626, Universit\'e Paris-Sud, 91405 Orsay, France}
\author{Robert L. Jaffe}
\affiliation{Center for Theoretical Physics, Laboratory for Nuclear Science, and Department of Physics, Massachusetts Institute of Technology, Cambridge, MA 02139, USA}
\begin{abstract}
We study the electromagnetic Casimir interaction of a compact object contained inside a closed cavity of another compact object. We express the interaction energy in terms of the objects' scattering matrices and translation matrices that relate the coordinate systems appropriate to each object. When the enclosing object is an otherwise empty metallic spherical shell, much larger than the internal object, and the two are sufficiently separated, the Casimir force can be expressed in terms of the static electric and magnetic multipole polarizabilities of the internal object, which is analogous to the Casimir-Polder result.  Although it is not a simple power law, the dependence of the force on the separation of the object from the containing sphere is a universal function of its displacement from the center of the sphere, independent of other details of the object's electromagnetic response. Furthermore, we compute the exact Casimir force between two metallic spheres contained one inside the other at arbitrary separations. Finally, we combine our results with earlier work on the Casimir force between two spheres to obtain data on the leading order correction to the Proximity Force Approximation for two metallic spheres both outside and within one another.
\end{abstract}
\maketitle

\section{Introduction}
Casimir forces arise due to vacuum fluctuations of electromagnetic fields in the presence of static or slowly moving conductors, or more generally, dielectric or magnetic materials~\cite{casimir}. The fields obey appropriate boundary conditions on the conductors or appropriate constitutive conditions on other electromagnetically active objects, which result in induced charges and currents. Due to the quantum nature of the field, the induced charges fluctuate, shifting the energy of the vacuum by a finite amount. This difference manifests itself as an interaction --- the Casimir force --- between neutral objects that depends on their sizes, shapes, material properties, and relative orientations. The case of perfect conductors is particularly simple: the Casimir force depends only on the geometry of the configuration. Analogous Casimir forces can arise from fluctuating scalar or fermion fields in the presence of objects on which they obey boundary or constitutive conditions. The electromagnetic Casimir force is a quantum effect observable at macroscopic scales. It has been shown to be significant in sub-micron scale devices as well as in the description of the interactions of atoms and/or molecules with surfaces, prompting substantial theoretical and experimental investigation over the last decade or so.

In this paper, we report computations of the force on a small polarizable object inside an otherwise empty conducting spherical shell as a function of its displacement from the shell's center --- the interior analog of the Casimir-Polder result. We further give the first exact calculation of the force between a metallic sphere inside a metallic spherical shell as a function of their radii and displacement. Although we restrict to metallic surfaces immersed in vacuum/air in this paper, the theoretical framework underlying our analysis is universal~\cite{Jamaluniversal}, and our methods have been extended to analogous interior configurations involving dielectric objects immersed in dielectric media~\cite{Jamal09}. Finally, we combine our results with earlier work on spheres~\cite{thorsten, thorstencyl} to obtain first order corrections to the Proximity Force Approximation (PFA) for two metallic spheres both outside and within one another. There has been much interest and research in computing the Casimir force beyond the PFA~\cite{beyondpfa, bordagpfa}; we answer this question in the case of perfectly conducting spheres. Some of the results described here were presented in an abbreviated form in an earlier Rapid Communication~\cite{Zaheerletter}.

In the past, there have not been many studies of the Casimir force in closed cavities despite the fact that cavity configurations are experimentally realizable. Marachevsky~\cite{marachevsky2001} computed the energy of a dilute dielectric sphere and a dipole at its center, and recently he studied the interaction of parallel plates inside a cylinder~\cite{plates}; Brevik {\it et al.}~\cite{brevik02} studied concentric dielectric spheres, and Dalvit {\it et al.}~\cite{cylinders} studied the interaction of a cylinder inside a cylinder. Recent theoretical advances~\cite{Jamaluniversal} (see also precursors Refs.~\cite{thorsten, KK, Maia_Neto}) in the study of the Casimir force have made it possible to analyze a wide variety of geometries and our investigation of the interior case is an example of configurations made accessible by these methods. In this paper, we shall only qualitatively summarize the path integral formalism to serve as a reminder for the reader. For an extensive introduction, a review of previous work, and further references we refer the reader to Ref.~\cite{Jamaluniversal}. 

The electromagnetic Casimir energy of an arbitrary configuration $\cC$ of objects can be calculated from the partition function $Z_{\cC}(\kappa)$, 
\begin{align}
\cE [\cC] &= -\frac{\hbar c}{2\pi} \int_{0}^{\infty} d\kappa \log \frac{Z_\cC(\kappa)}{Z_{0} (\kappa)}, \label{eq:general}
\end{align}
where $Z_{\cC}(\kappa)$ is obtained from the Minkowski space functional integral $\cZ(T)= \int \mathcal{D}\mathbf{A} e^{\frac{i}{\hbar}S[T]}$ ($S[T]$ is the electromagnetic action evaluated from $t=0$ to $t=T$.) after decomposing the fields $\bE$ into their fourier modes and Wick-rotating to imaginary time--- $\kappa$ being the imaginary frequency. The restriction $\kappa \geq 0 $ allows for $\bE(ic\kappa)$ and $\bE^*(ic\kappa)$ to be considered independently since $\bE$ is real. We subtract the Casimir energy at a convenient location (described by $Z_0$ in the denominator of the $\log$) to remove the cut-off dependent terms in the unrenormalized energy, since such contributions arise from the objects individually.

The spatial configuration of interacting objects manifests itself physically as a continuity condition that the fluctuating field $\bE$ obeys on their surfaces. It is possible to trade the constraints on $\bE$ for fluctuating sources~\cite{Jamaluniversal, thorsten,kardar}. Then the functional integral over the fields becomes free of constraints and can be performed up to a multiplicative constant. Since the path integral over the free fields is independent of the location of the objects in space, the multiplicative constant is cancelled when dividing by the partition function for the reference configuration. This leaves a functional integral over the fluctuating sources on each object $\alpha$, in which the action is expressed as a functional of the sources and the classical field $\bE_{\text{cl}}$ they produce. By superposition, we write $\bE_{\text{cl}} = \sum_\alpha \bE_{\alpha, \text{cl}}$, which allows the action to be written as a sum over the self- and inter-actions of all the objects in the system. 

With the choice of convenient bases, we can expand $\bE_{\alpha, \text{cl}}$ in terms of multipole fields, generated by the multipole moments of the sources induced on $\Sigma_\alpha$. We can express these multipole fields in terms of the transition matrix $\cT = (\mathcal{S} - \mathcal{I})/2$ of the object under consideration (where $\mathcal{S}$ is its scattering matrix) and the multipole sources. The self-action of the sources on each object can, therefore, be written entirely in terms of the multipole moments and its $\cT$-matrix. Similarly, we can express the inter-action of two different objects in terms of their multipoles and a translation matrix which relates their coordinate systems in the appropriate bases. Finally, the functional integral over the multipole moments of the sources can be performed, leaving an expression for the Casimir energy in terms of the objects' $\cT$-matrices and the translation matrices.

We are interested in a situation where one object, the {\it internal} (subscript $i$), is enclosed entirely within a cavity of another, the {\it external} object (subscript $e$). The polarizable internal object interacts with fields scattered {\it inside} the cavity in which it is immersed. Therefore the $\cT$-matrix of the external object relevant to the interior case differs from the $\cT$-matrix that describes scattered waves outside the external object. Following Ref.~\cite{Jamaluniversal}, we denote the $\cT$-matrices of the internal and external objects by $\cT^{ee}_{i}$ and $\cT^{ii}_{e}$ respectively, where the subscript denotes the object and the superscripts denote the relevant scattering amplitude ($\cT^{ee}$ being the standard $\cT$-matrix of an object). For conducting boundary conditions on the cavity of the external object, $\cT^{ii}_e = [\cT^{ee}_e]^{-1}$, where $\cT^{ee}_e$ would be the standard $\cT$-matrix for scattering exterior to a conductor in the shape of the cavity.  Additionally, the translation matrices in the interior problem are different ($\mathcal{V}$ instead of $\mathcal{U}$ as employed in Refs.~\cite{thorsten,scalar}) because we are interested in quantum fluctuations internal to the cavity and external to the internal object. The $\mathcal{V}$-matrices appear because they relate regular waves to outgoing waves as opposed to the $\mathcal{U}$-matrices that relate outgoing waves to outgoing waves. With these modifications, Eq.~(\ref{eq:general}) evaluates to  
\begin{align}
\mathcal{E}&= \frac{\hbar c}{2 \pi } \int_0^{\infty} d\kappa \ln \frac{\det (\mathcal{I} -  \cT_{e}^{ii}  \mathcal{V}_{e,i}  \mathcal{T}_{i}^{ee}  \mathcal{V}_{i, e})}{\det (\mathcal{I}- \mathcal{T}_e^{ii}  \mathcal{T}_i^{ee})}. \label{eq:master}
\end{align}
The dielectric properties of the two objects, and the medium separating them inside the cavity are encoded in the respective $\cT$ and $\mathcal{V}$-matrices. The denominator subtracts the energy when the centers of the two objects coincide (as opposed to infinitely separated in an exterior problem~\cite{thorsten,scalar}). In this way we eliminate the cutoff dependent Casimir energy of the dipole at the center of the sphere~\cite{marachevsky2001}. For a detailed discussion, we refer the reader to Ref.~\cite{Jamaluniversal}.

Eq.~(\ref{eq:master}) can be evaluated exactly for certain geometries for which the $\cT$ and $\mathcal{V}$-matrices are easily calculable in a convenient basis. The case of spherically symmetric dielectric objects is particularly simple because their $\cT$-matrices are diagonal in the basis of spherical wave functions. In this paper, we provide results for a metallic sphere inside an otherwise empty metallic spherical shell, while dielectric objects immersed in a dielectric medium are treated in Ref.~\cite{Jamal09}.

The matrix identity $\ln \det \mathbb{M} = \text{Tr} \ln\mathbb{M}$, allows for a simple physical interpretation of Eq.~(\ref{eq:master}). We can express the Casimir energy as a series, ${\cal E}= \hbar c/2\pi  \int d\kappa{\rm Tr}\left( \mathcal{N} + \frac{1}{2}\mathcal{N}^{2}+...\right)$, over the matrix $\mathcal{N} = \mathcal{T}_{e}^{ii}  \mathcal{V}_{e,i}  \mathcal{T}_{i}^{ee}  \mathcal{V}_{i, e}$ where $\mathcal{N}$ describes a wave that travels from one object to the other and back~\cite{thorsten}. In general, all terms in this series are important, illustrating the fundamentally non-two-body nature of the Casimir force. The rate of convergence of this series depends on the size of the internal object relative to the separation of its surface from that of the cavity. 

First we consider an object that is small compared to the size of the cavity. Then the first term in the series expansion of Eq.~(\ref{eq:master}), ${\cal E}= \hbar c/2\pi  \int d\kappa{\rm Tr}\mathcal{N}$, already gives an excellent approximation to the energy. Furthermore, in this limit the Casimir energy is dominated by the lowest frequency contributions from the lowest partial waves in $\mathcal{T}_i^{ee}$. In a spherical basis, the leading terms in the electromagnetic $\mathcal{T}$-matrix are, $\mathcal{T}^{\lambda \lambda}_{lml'm'} \sim \kappa^{l+l'+1}$ and $\mathcal{T}^{\lambda \sigma}_{lml'm'} \sim \kappa^{l+l'+2}$ for $\lambda \not = \sigma$, where $l=1,2,...$, and $\lambda$ and $\sigma$ label the polarizations $E$ (electric) or $M$ (magnetic). The leading contribution to the Casimir force comes from the orientation dependent dipole response of the internal object to a dipole field, where the internal object can be characterized by its polarizability tensor, $\alpha^{M/E}_{mm'} = \frac{3}{2}\kappa^{-3}\mathcal{T}^{M/E}_{1m1m'}$ (see Ref.~\cite{orientation}). 

Now we fix the external object to be a conducting spherical shell of radius $R$ and define $a$ to be the displacement of the center of the internal object from the center of the shell. To leading order in $r/R$ (where $r$ is the typical size of the internal object), the Casimir energy can be expressed as~\cite{Zaheerletter},
\begin{align}
  &\frac{3\pi R^{4}}{\hbar c} \mathcal{E}(a/R)     = \left[f^{E}(a /R)-f^E(0)\right]  {\text{Tr} \alpha^E}  + g^{E}(a/R) (2 \alpha^E_{zz}-\alpha^E_{xx}- \alpha^E_{yy})   + (E \leftrightarrow M) + \ldots \label{eq:orientation0}
\end{align}
where the polarizability tensor $\alpha_{mm'} \sim r^3$ has been expressed in a Cartesian basis, and ``$+...$'' denotes terms that are higher order in $r/R$. The functions $f$ and $g$ are plotted in Fig.~\ref{hdme} and their functional forms are given in Section~\ref{casimirpolder}. Eq.~(\ref{eq:orientation0}) describes the interaction of a polarizable object (an atom for instance) inside a conducting spherical shell, which is analogous to the well-known Casimir-Polder potential~\cite{casimirpolder}. However it differs from the Casimir-Polder result in three ways: $f,g$ are non-trivial functions of $a/R$; the internal object experiences a torque; and the Casimir force between the two objects depends on the internal object's orientation. Note that the expansion in Eq.~(\ref{eq:orientation0}) is asymptotic in $r/R$ at fixed $a/R$. For a spherically symmetric internal object, the orientation dependent terms in the Casimir energy vanish, and corrections to Eq.~(\ref{eq:orientation0}) come from the static quadrupole electric and magnetic polarizabilities, $\alpha_{2}^{M,E}$ of the internal object. These corrections are given in the Appendix. 

The opposite extreme from the Casimir-Polder limit occurs when the interior object nearly touches the cavity wall.  The leading behavior of the Casimir force in this limit is given by the Proximity Force Approximation~\cite{PFA}. The PFA prediction for the Casimir {\it force} between two conducting spheres, whether they are separated or contained one inside the other is given by
\begin{align}
\lim_{d\to 0}d^{3}\,{\cal F}(d,s,R) = -\frac{\pi^3 \hbar c}{360} \frac{rR}{R+r}  \quad , \label{pfalimit}
\end{align}
where $r$ and $|R|$ are the radii of the internal and external spheres respectively and $d$ is the minimum distance between their surfaces.  By convention we keep $r$ fixed and let $R$ vary.  $R>0$ corresponds to the {\it exterior} problem (the spheres are separated); $R<0$ corresponds to the {\it interior} problem; in the limit $R\to\infty$ we have a sphere opposite a plane. The constraint $r\leq|R|$ avoids double counting of sphere-sphere configurations. Eq.~(\ref{pfalimit}) is derived for $R>0$ by semi-classical methods in Ref.~\cite{Schaden:1998zz} and the extension to $R<0$ is straightforward but its corrections have up to now not been known. The planar and exterior problems have been studied in Ref.~\cite{thorsten} and Ref.~\cite{thorstencyl}, respectively (see also Refs.~\cite{beyondpfa, bordagpfa}). Since most experiments up to now have considered spherical conductors separated by distances much smaller than their radii, the first correction in $d/r$ to the PFA is the geometric correction of greatest immediate interest.  As discussed in 
Section~\ref{pfa}, we parameterize the first correction to the PFA by
\begin{align}
\mathcal{F} (d,r,R) = -\frac{\pi^3 \hbar c}{360d^3} \frac{Rr}{R+r}\left(1 + \theta_{1}(r/R) \frac{d}{2r} - \theta_2(r/R) \frac{d^2}{2r^2}  + \cO(d^3/r^3) \right). 
 \label{eq:pfainside}
\end{align}
Our evaluation of Eq.~(\ref{eq:master}) for conducting spheres has allowed us to combine our results with those of Refs.~\cite{thorsten,thorstencyl} to predict an estimate of the PFA correction coefficient $\theta_1(r/R)$ appearing in Eq.~(\ref{eq:pfainside}) for $-1\le r/R \le 1$. We refer the reader to Section~\ref{pfa} for further discussion.

The rest of this paper is organized as follows: Section~\ref{vector} provides representations of the vector transition and translation matrices relevant to the interior case in a spherical wave basis, followed by an exact computation of the Casimir force between metallic spheres in Section~\ref{numeric}.  In Section~\ref{pfa}, we discuss first order corrections  to the Proximity Force Approximation for two metallic spheres of arbitrary size based on numerical results in this paper and in~\cite{thorsten, thorstencyl}. In Section~\ref{casimirpolder}, we derive the interior Casimir-Polder result, and study its comparison with the exact results of Section~ \ref{numeric}. Ref.~\cite{thesis} repeats the analysis of this paper for a complex scalar field, which follows by analogy with the vector case.

\begin{figure}[t]
\includegraphics[width=8cm]{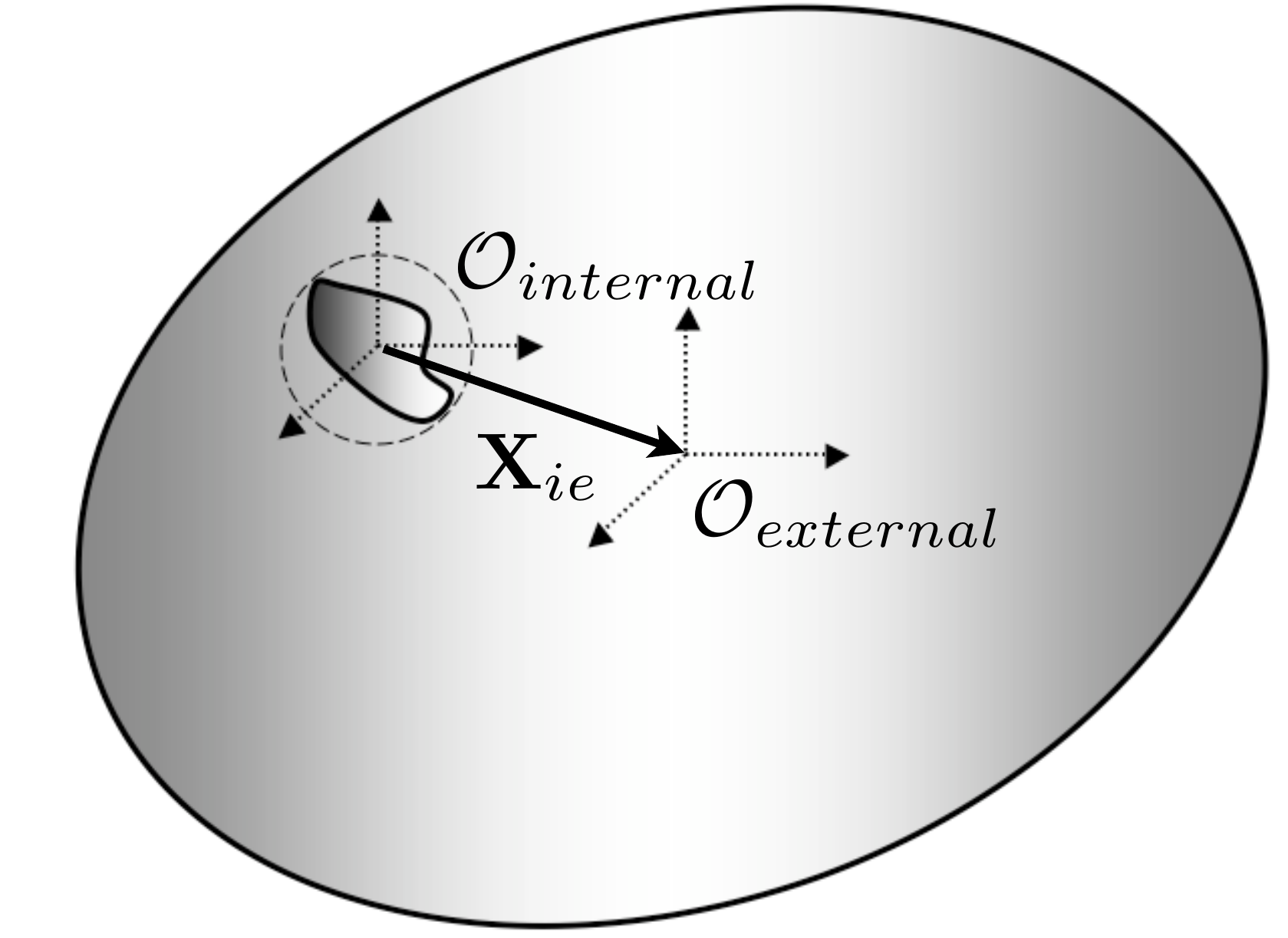}\caption{
Interior Geometry: An object inside the cavity of an external object. We assume that it is possible to choose a bounding sphere that does not overlap with the surface of the cavity. The distance between the two origins is denoted by $\bf{X}_{ie}$. \label{interior_diagram}}
\end{figure}

\section{Theoretical background}\label{vector}
As pointed out in the introduction, the Casimir energy of a configuration of compact objects can be evaluated in terms of their transition matrices $\cT= (\mathcal{S}-\mathcal{I})/2$ (where $\mathcal{S}$ is the scattering matrix) and translation matrices that relate various coordinate systems relevant to each object. For the interior geometry, we describe the internal object by $\cT_{i}^{ee}$, the scattering amplitudes for scattering outside its external surface, and the external object by $\cT_{e}^{ii}$, the scattering amplitudes for scattering \emph{inside} the cavity of the external object which contains the internal object. We use the subscript $M$ for the medium inside the cavity, and index the dielectric constants $\epsilon_x(ic\kappa)$ and $\mu_x(ic\kappa)$ by $x = (i, e, M)$. Fig.~(\ref{interior_diagram}) is an illustration of the interior geometry. 

Once the geometry and the dielectric properties of the interacting objects are specified, the next step is to calculate the $\cT$ and $\mathcal{V}$-matrices in a convenient basis. We specialize to spherical coordinates and solve the interior problem exactly for a sphere contained inside a spherical cavity. Choosing $r$ and $R$ to denote the radii of the internal object and the cavity respectively, the $\cT$-matrices are diagonal, and are represented in a spherical basis as,
\begin{align}
\cT_{i, lmMl'm'M}^{ee} &= -\delta_{ll'} \delta_{mm'}\frac{\mu_Mi_l(n_M\kappa r)\partial_{r}(ri_l(n_i\kappa r))- \mu_i\partial_r(ri_l(n_M\kappa r))i_l(n_i\kappa r)}{\mu_Mk_l(n_M\kappa r)\partial_{r}(ri_l(n_i\kappa r))- \mu_i\partial_r(rk_l(n_M\kappa r))i_l(n_i\kappa r)}   \nonumber \\
&\xrightarrow[\epsilon_{i}\to\infty]{} -\delta_{ll'} \delta_{mm'} \frac{i_l(n_M\kappa r)}{k_l(n_M\kappa r)}\nonumber \\
\cT_{i, lmEl'm'E}^{ee} &= -\delta_{ll'}\delta_{mm'}\frac{\epsilon_Mi_l(n_M\kappa r)\partial_{r}(ri_l(n_i\kappa r))- \epsilon_i\partial_r(ri_l(n_M\kappa r))i_l(n_i\kappa r)}{\epsilon_Mk_l(n_M\kappa r)\partial_{r}(ri_l(n_i\kappa r))- \epsilon_i\partial_r(rk_l(n_M\kappa r))i_l(n_i\kappa r)} \nonumber \\
&\xrightarrow[\epsilon_{i}\to\infty]{}-\delta_{ll'}\delta_{mm'}\frac{\partial_r(ri_l(n_M\kappa r))}{\partial_r(rk_l(n_M\kappa r))} \label{eq:tmatrix}
\end{align}
where $n_x= \sqrt{\epsilon_x\mu_x}$ is the refractive index, and $M,E$ denote the magnetic and electric polarizations respectively. $\cT_{e}^{ii}$ can be obtained by the substitutions $i_l \leftrightarrow k_l$, $r\to R$, and the subscript $i\to e$ for the dielectric constants everywhere in Eqs.~(\ref{eq:tmatrix}). $k_l$ and $i_l$ are modified spherical bessel functions of integer index $l$. 

The spherical wave translation matrices that relate the coordinate systems centered on the internal sphere and the external cavity are given by~\cite{Jamaluniversal, wittmann}, 
\begin{align}
\mathcal{V}_{ie, l'm'M,lmM} &= (-1)^m \sum_{l''} [l(l+1)+l'(l'+1)-l''(l''+1)] \sqrt{\frac{\pi(2l+1)(2l'+1)(2l''+1)}{l(l+1)l'(l'+1)}}\nonumber \\
& \times \left(\begin{array}{ccc} l & l'& l''\\ 0&0&0\end{array}\right)\left(\begin{array}{ccc} l & l'& l''\\ m&-m'&m'-m\end{array}\right) i_{l''}(n_M\kappa|\mathbf{X}_{ie}|)(-1)^{l''} Y_{l'' m-m'}(\hat{\mathbf{X}}_{ie})\nonumber 
\end{align}
\begin{align}
\mathcal{V}_{ie, l'm'E,lmM} &= - \frac{in_M\kappa}{\sqrt{l(l+1)l'(l'+1)}} \mathbf{X}_{ie}\add{\cdot} \Big[\hat{\bx} \frac{1}{2} \left(  \lambda^+_{lm} B_{l'm'lm+1}(\mathbf{X}_{ie}) + \lambda^-_{lm}B_{l'm'lm-1}(\mathbf{X}_{ie})  \right)  \, ,  \nonumber \\
&+ \hat{\mathbf{y}}\frac{1}{2i} \left( \lambda^+_{lm} B_{l'm'lm+1}(\mathbf{X}_{ie})- \lambda^-_{lm}B_{l'm'lm-1}(\mathbf{X}_{ie}) \right)  + \hat{\mathbf{z}}mB_{l'm'lm}(\mathbf{X}_{ie}) \Big] ,\nonumber \\
 \mathcal{V}_{ie, l'm'M,lmE} &= -\mathcal{V}_{ie, l'm'E,lmM}  , \qquad
 \mathcal{V}_{ie, l'm'E,lmE} = \mathcal{V}_{ie, l'm'M,lmM} \, , \label{eq:vmatrix}
\end{align}
where 
\begin{align}
B_{l'm'lm}(\mathbf{X})= (-1)^m \sum_{l''} \sqrt{4\pi(2l+1)(2l'+1)(2l''+1)}\left(\begin{array}{ccc} l & l'& l''\\ 0&0&0\end{array}\right)\left(\begin{array}{ccc} l & l'& l''\\ m&-m'&m'-m\end{array}\right) i_{l''}(n_M\kappa|\mathbf{X}|)(-1)^{l''} Y_{l'' m-m'}(\hat{\mathbf{X}}) \nonumber
\end{align}
and $\lambda^{\pm}_{lm} = \sqrt{(l\mp m)(l\pm m+1)}$. $\mathbf{X}_{ie}$ is the displacement vector that extends from the center of the internal object to the center of the cavity with $a=|\mathbf{X}_{ie}|$. The translation matrix $\mathcal{V}_{ei}$ is related to $\mathcal{V}_{ie}$ by,
\begin{align}
\mathcal{V}_{ei} = \left(\begin{array}{cc} 1&0\\0&-1 \end{array}\right)\mathcal{V}_{ie}^\dagger \left(\begin{array}{cc} 1&0\\0&-1 \end{array}\right)
\end{align}
in the two dimensional space of electric ($E$) and magnetic ($M$) polarizations. To simplify calculations, we have aligned the $\mathbf{z}$-axis of the cavity along $\mathbf{X}_{ie}$. The Casimir energy of dielectric spheres immersed in a dielectric medium is derived using the above equations in Ref.~\cite{Jamal09}. We present results for conducting boundary conditions in the following section.

\section{Computation for metallic spheres}\label{numeric}
\begin{figure}[t]
\includegraphics[scale=0.5]{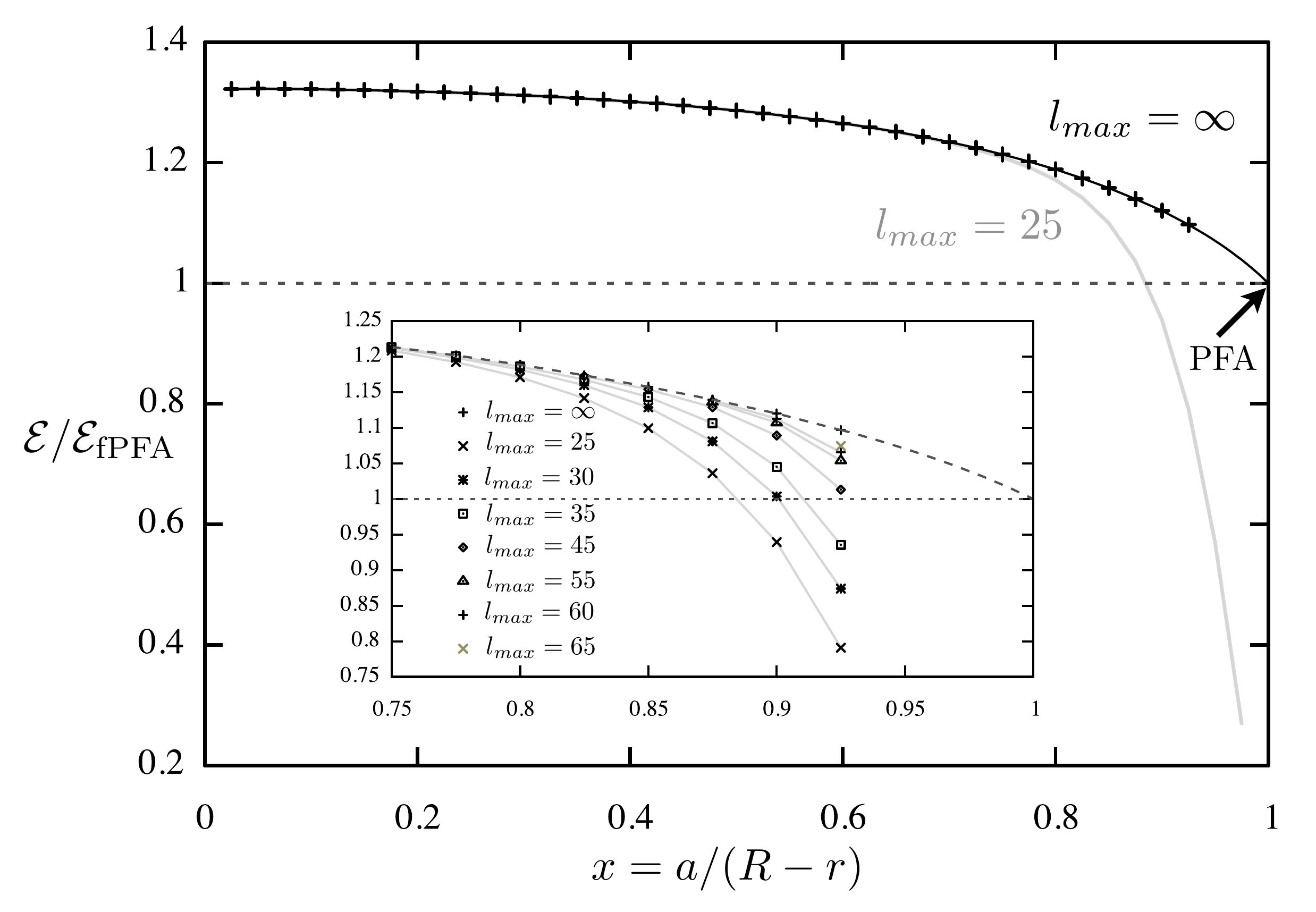} \caption[Short]{Casimir energy between two conducting spheres. The dark black line shows the Casimir energy, $R(x) = \cE/\cE_{\rm fPFA}$, as a function of $x = a/(R-r)$ where $a$ is the displacement of centers. The radius of the inner sphere is fixed at $r = 0.5R$, where $R$ is the radius of the outer sphere. In the limit $x\to 1$, the Casimir energy approaches the PFA energy, which is marked by the dashed horizontal line. $\mathcal{E}_{\rm fPFA}$ denotes the `full' form of the PFA energy discussed in Section~\ref{pfa}. At intermediate separations, the Casimir energy is dominated by lower partial waves. For example, the gray line shows that the energy obtained by integrating Eq.~(\ref{eq:master}) to partial wave order $l=25$ is accurate up to $x \sim 0.7$. The dark line  is obtained by extrapolating to $l= \infty$. Inset: Convergence at close separations, $0.75 \leq x\leq 1$. \label{energy}}
\end{figure}

In this section, we analyze the Casimir interactions of a metallic sphere contained within an otherwise empty metallic spherical shell. The Casimir energy is obtained by numerical integration of Eq.~(\ref{eq:master}) using the matrix representations given in Section~\ref{vector}. It depends on the ratio of the radii, $r/R$, and varies with the displacement $a$ of the centers, parameterized by $x=a/(R-r)$. $x$ ranges from zero (the spheres are concentric) to unity (the two spheres touch). The Casimir energy varies over many orders of magnitude and indeed diverges as the spheres touch at $x=1$.  To make our graphs easier to read and the subsequent numerical fits easier to perform, we seek to divide the Casimir energy by a simple function that captures the growth near the limit $x\to 1$.  This can be accomplished by utilizing the Proximity Force Approximation~\cite{PFA}, which accurately predicts that in the limit $x\to 1$, the Casimir energy diverges as $(1-x)^{-2}$.  Unfortunately, the leading term in the PFA does not behave correctly as $x\to 0$ where the force should vanish. (The Casimir energy is quadratic in $x$ for small $x$.)   To accommodate both limits, $x\to 0$ and $x\to 1$, we employ an extension of the PFA which although has no theoretical foundation (beyond the leading term) yet provides a definite and convenient function that captures the dominant variation in the exact Casimir energy over the whole range of $x$.  We refer to this function as the ``full PFA'', and denote it by $\cE_{\rm fPFA}(x)$; it is described in detail in  Section~\ref{pfa} (see Eq.~(\ref{eq:fullpfa})). To illustrate our results, we choose $r/R= 0.5$ and plot $R(x) = \cE(x)/\cE_{\rm fPFA}(x)$ in Fig.~\ref{energy}.

Given data like that shown in Fig.~\ref{energy}, we obtain the Casimir \emph{force} by numerically differentiating $R(x)$ with the help of the relation,  
\begin{equation}
\label{eq:forceenergy}
\frac{\mathcal{F}(x)}{\mathcal{F}_{\rm fPFA}(x)} = R(x) + \frac{\mathcal{E}_{\rm fPFA}(x)}{\mathcal{F}_{\rm fPFA}(x)} R'(x).
\end{equation}
Note that $R(x)$ is determined numerically only up to $x=0.925$. Therefore in the range $x \leq 0.9$, $R'(x)$ is determined by numerical differentiation. For $x \geq 0.9$, $R'(x)$ is determined either by differentiating a suitable function that extrapolates $R(x)$ or by extrapolating $R'(x)$ itself. Both procedures give identical results (within numerical error, as discussed below). The numerical integration and differentiation were performed with MATLAB while all fitting and extrapolation procedures were performed with GNUPLOT.

Although all partial waves contribute to the Casimir energy, partial waves with $l\leq l_{\rm max}$ contribute the most; where $l_{\rm max}$ depends on the spheres' relative sizes and separation and grows rapidly as the separation gets small. For $r/R =0.5$, the gray curve in Fig.~\ref{energy} shows the results for $l_{\text{max}} \leq 25$. For $x > 0.7$ the limitation to $l_{\rm max}< 25$ is inadequate.  To obtain results in this range of $x$ it is necessary to include progressively larger values of $l$. Eventually the numerical evaluation of Eq.~(\ref{eq:master}) is limited by our ability to manipulate large matrices. For example $l_{\text{max}} > 65$ at $x\sim 0.9$. To obtain accurate results for $x>0.7$ we first compute ${\cal E}(l_{\rm max})$ for a sequence of values of $l_{\rm max}$.  Then we fit ${\cal E}(l_{\rm max})$ to a decaying exponential (which seems to capture the leading behavior at large $l_{\rm max}$) of the form $\cE(l_{\rm max}) = \cE(\infty) - \alpha e^{-\beta l_{\rm max}}$, where $\alpha$ and $\beta$ are constants.  The resulting function, ${\cal E}(\infty)$ is plotted in Fig.~\ref{energy}.  It smoothly extrapolates to the PFA point at $x=1$ as it should.  The convergence of ${\cal E}(l_{\rm max})$ to ${\cal E}(\infty)$ is illustrated in Fig.~\ref{energy} (inset).

At even closer separations, $x \geq 0.925$, important contributions to the Casimir energy come from values of $l$ even larger than $l\approx 65$. It is difficult to evaluate numerically stable values for matrix elements involving modified spherical bessel functions $k_l(x)$ in the limit $x \to 0$ with $l\sim 65$ and above. Therefore our numerical methods are inadequate when $l$ grows above 65. But we know that the Casimir energy approaches the PFA limit as $x\to 1$. Therefore we extrapolate the exact data calculated at $x\leq 0.925$ to estimate the Casimir energy for $0.925 \leq x\leq 1$. For example, for the case shown in Fig.~\ref{energy}, this is achieved by extrapolating the five data points between $0.825\leq x \leq 0.925$ to a function, $f(d/r) = 1+ \bar{\theta}_1d/r +\bar{\theta}_2\log(d/r) d^2/r^2 $, where $d = R-r-a$, $\bar{\theta}_1 = 1.770 \pm 0.034$ and $\bar{\theta}_2 = 2.272 \pm 0.271$. Notice that $\bar{\theta}_{1} = \theta_{1}- \theta_{1,\text{fPFA}}$ where $\theta_{1}$ is defined for all $r/R$ in Eq.~(\ref{eq:pfainside}) and $\theta_{1, \text{fPFA}}$ is calculated from Eq.~(\ref{eq:fullpfa}), and likewise for $\bar\theta_{2}={\theta_2}-\theta_{2,\rm{fPFA} }$.  Therefore, the values of the PFA correction coefficients for $r/R =0.5$ can be easily determined. The above analysis can be easily repeated for other values of $r/R$ and the coefficients $\theta_1$ and $\theta_{2}$ determined for a range of those values. Thus our numerical methods yield sub-leading corrections to the PFA. We perform these computations in Section~\ref{pfa}. For more details, the reader is referred to that Section.
\begin{figure}[t]
\includegraphics[scale=0.5]{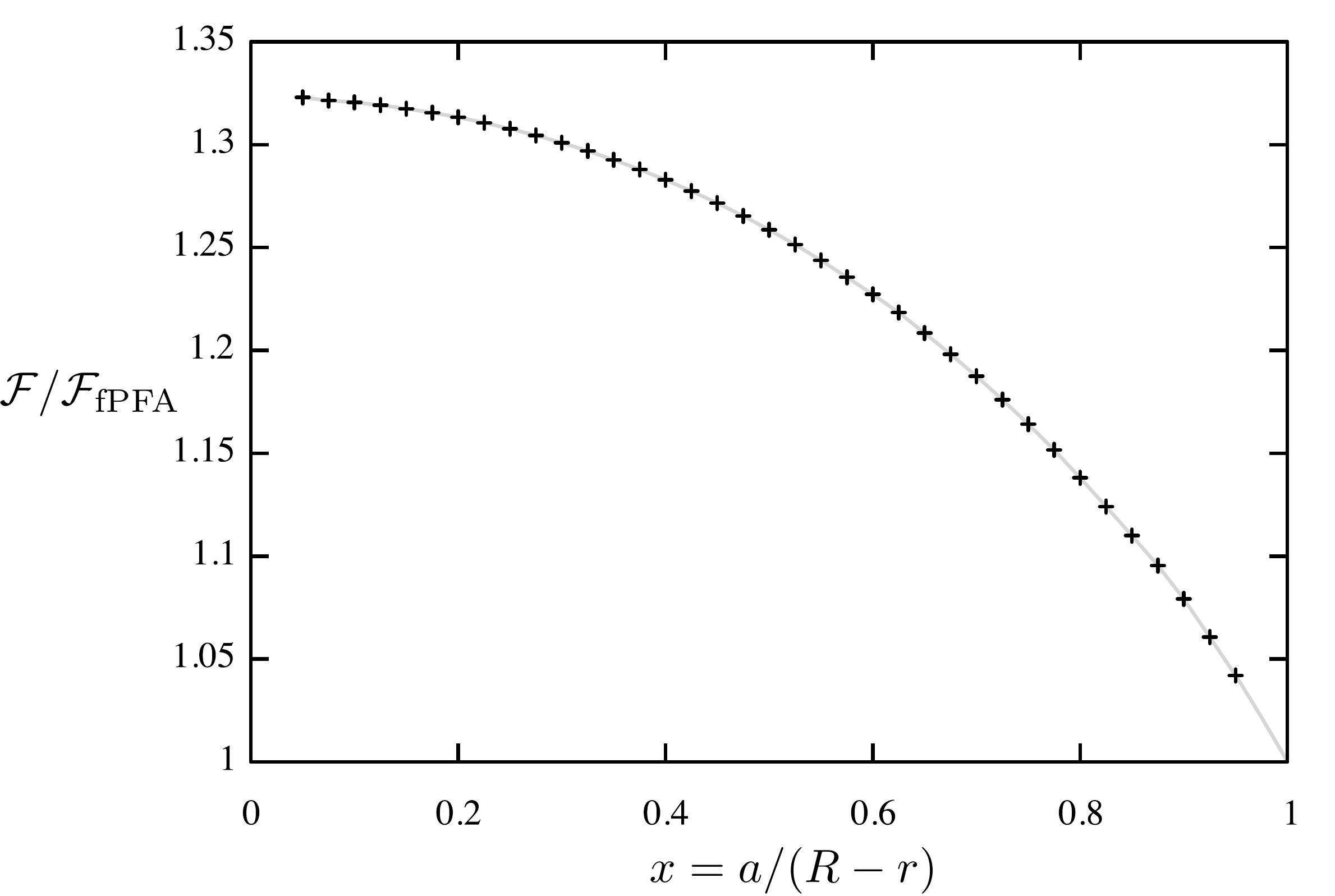} \caption[Short]{Casimir force between two conducting spheres. The black line shows the Casimir force, $\mathcal{F}/\mathcal{F}_{fPFA}$, between two conducting spheres as a function of $x= a/(R-r)$ where $a$ is the displacement of their centers. The radius of the inner sphere is fixed at $r = 0.5R$, where $R$ is the radius of the outer sphere. In the limit $x\to 1$, the Casimir force approaches the PFA. $\mathcal{F}_{\rm fPFA}$ denotes the `full' form of the PFA discussed in Section~\ref{pfa}. \label{force}}
\end{figure}

The Casimir force between two conducting spheres depicted in Fig.~\ref{force} is calculated by numerical differentiation of the exact data points spaced at $\Delta x =0.025$ along the black curve $R(x)$ in Fig.~\ref{energy}. We remind the reader that Fig.~\ref{energy} plots $R(x) = \cE/\cE_{\rm fPFA}$ as a function of $x$. The curve shown  in Fig.~\ref{force} is calculated from $R(x)$ using Eq.~(\ref{eq:forceenergy}).
The differentiation of $R(x)$ is performed using centered differences for 37 data points between 0.05 and 0.925. For $x\geq 0.9$, $\mathcal{F}/\mathcal{F}_{\rm fPFA}$ can be determined either by an independent extrapolation or by an algebraic manipulation of the fit describing $R(x)$ in that range. We fit a function of the form, $h(d/r) = 1 + \bar{\theta}_1 d/2r - \bar{\theta}_2d^2/2r^2- \theta_{1,\text{fPFA}}(\bar{\theta}_1+ \theta_{1,\text{fPFA}}) d^2/4r^2$ to the four data points between $x=0.825$, and $x=0.9$ and compare the new extrapolation constants with the ones determined in Fig.~\ref{energy}. The function $h$ is determined algebraically from $\mathcal{F}/\mathcal{F}_{\rm PFA}$, where $\mathcal{F}$ is given by Eq.~(\ref{eq:pfainside}) We find $\bar{\theta}_1= 1.770 \pm 0.032$ and $\bar{\theta}_2 = 2.058 \pm 0.529$, which agree with their previously calculated values within their error. This achieves two goals: it makes contact with the PFA prediction in Eq.~(\ref{eq:pfainside}), and demonstrates  that the function $f(d/r)$ used to extrapolate $R(x)$ at values of $x$ close to 1 was the correct {\it ansatz} for the sub-leading PFA behavior for the energy in Fig.~\ref{energy}. 

We have illustrated the numerical evaluation of Eq.~(\ref{eq:master}) with the case $r/R=0.5$. However, the same techniques may be applied to determine the Casimir force and energy by numerically integrating Eq.~(\ref{eq:master}) for all configurations, $0<r/R<1$. In the following Section, we apply the methods demonstrated in this section to study various $r/R$ configurations in the limit $x \to 1$ and determine the PFA correction coefficients $\theta_1$ corresponding to those configurations. On the other hand, the Casimir force for intermediate values of $a/R$ is studied completely analytically for the range of interior configurations $r/R \to 0$ in Section~\ref{casimirpolder}.

\section{Corrections to the PFA}\label{pfa}
As mentioned in the Introduction, one of the most interesting quantities made accessible by our methods is the first non-trivial correction to the Proximity Force Approximation. In this section we extract this correction for the case of one sphere within another, and combine it with data from the cases of two separated spheres and a sphere opposite a plane, to survey the full range of possible sphere-sphere configurations.

The leading term in the PFA is given by Eq.~(\ref{pfalimit}) as discussed in the Introduction. The analytic form of the corrections to the PFA is unknown in general, however the case of a sphere facing a plane was treated analytically in Ref.~\cite{bordagpfa}. We find that our data can be fitted very well by the first few terms in a {\it power series expansion of the Casimir force} in $d/r$, as quoted in the Introduction, 
\begin{align}
 \mathcal{F}(d,r,R)= -\frac{\pi^3 \hbar c}{360 d^3 } \frac{rR}{R+r} \left( 1 + \theta_{1}(r/R)\frac{d}{2r} - \theta_{2}(r/R)\frac{d^{2}}{2r^{2}}+ \cO(d^3/r^3) \right)\label{eq:genexpansion}
\end{align}
as $d\to 0$. (Remember that $r/R<(>)0$ corresponds to the interior (exterior) case.) This power series expansion of the force requires that the energy include a $\log(d/r)$ term,
\begin{align}
 \mathcal{E}(d,r,R)= -\frac{\pi^3 \hbar c}{720 d^2 } \frac{rR}{R+r} \left( 1 + \theta_{1}(r/R)\frac{d}{r} + \theta_{2} (r/R) \log(d/r){\frac{d^{2}}{r^{2}}}+ \alpha(r/R)\frac{d^{2}}{r^{2}} +\cO(d^3/r^3)\right) \label{eq:pfa}
\end{align}
where we have adjusted signs in Eq.~(\ref{eq:genexpansion}) so that corrections to the PFA energy in Eq.~(\ref{eq:pfa}) correspond to the extrapolation function $f(d/r)$ defined in Section~\ref{numeric}. Note that the term proportional to $d^{2}/r^{2}$ in Eq.~(\ref{eq:pfa}) does not contribute to the force. 

It is useful to have an estimate, however crude, of the interior Casimir energy over the whole range of $d/R$ in order to scale out the rapid variation that makes it difficult to display ${\cal E}$ graphically.  To this end we extend the PFA over the whole range of $d$, $r$, and $R$. The PFA estimate of ${\cal E}$ can be calculated by assuming that each interacting surface is assembled out of infinitesimal mirrors spaced at a distance $l(\zeta_1,\zeta_2)$ from the other surface, where $(\zeta_1, \zeta_2)$ are the coordinates of the surface chosen as a convenient reference. This algorithm is ambiguous beyond the leading term in $1/d$ because there is no unique way to specify the separation between the surfaces.  For definiteness, we extend the PFA by taking $d$ to be the distance between the surfaces measured radially outward from the smaller sphere and integrate over the surface of the smaller sphere.  This can be done for both the interior ($r/R<0$)and the exterior ($r/R>0$) configurations. Note that the restriction $y=r/R \in [-1,1]$ covers the full range of sphere-sphere configurations as long as $r$ is taken to be the radius of the smaller sphere. The result, which we refer to as the `full PFA', is given here (for compactness) as a definite integral (which can also be obtained in closed form), 
\begin{align}
\cE_{\text{fPFA}}   =& -\frac{\pi^3\hbar c y^2}{360R} \int_{-1}^1 dx \left(\frac{1}{\left(-(1+y+yd/r)x+y+\sqrt{(1+y+yd/r)^2(x^2 -1)+1}\right)^3}+\frac{1}{(1+y)^3}\right), &\quad y <0 \nonumber 
\end{align}
\begin{align}
\cE_{\text{fPFA}}   =& -\frac{\pi^3\hbar c y^2}{360R} \int_{x_{0}(y,d/r)}^1 \frac{dx}{\left((1+y+yd/r)x-y-\sqrt{(1+y+yd/r)^2(x^2 -1)+1}\right)^3},&\quad y>0\nonumber \\
&\mbox{where}\,\,x_{0}(y,d/r)={\sqrt{1-\frac{1}{(1+y+yd/r)^2}}}.
\label{eq:fullpfa}
\end{align} 
(For $y <0$ we have subtracted the energy when the spheres are concentric as in Eq.~(\ref{eq:master}).) 

Another option would be to measure the distance $d$ radially {\it inward} from the outer sphere. This procedure yields a different analytical form for the ``full PFA'' beyond the leading order, illustrating the ambiguity in defining $d$. To distinguish between the two ``full PFA'' estimates, we refer to the former as $r$-based and the latter as $R$-based.

If we expand around $d/r=0$ in Eq.~(\ref{eq:fullpfa}), we find that $\theta_{1, \rm fPFA} (r/R) = -r/R - r/(R+r)-3$, which is continuous in the interval $r/R \in (-1,1]$. A similar calculation from the $R$-based PFA yields a different, yet continuous, form for $\theta_{1, \text{fPFA}}$, i.e. $-(3r/R + r/(r+R)+1)$. 

\begin{figure}[t]
\includegraphics[scale=0.4]{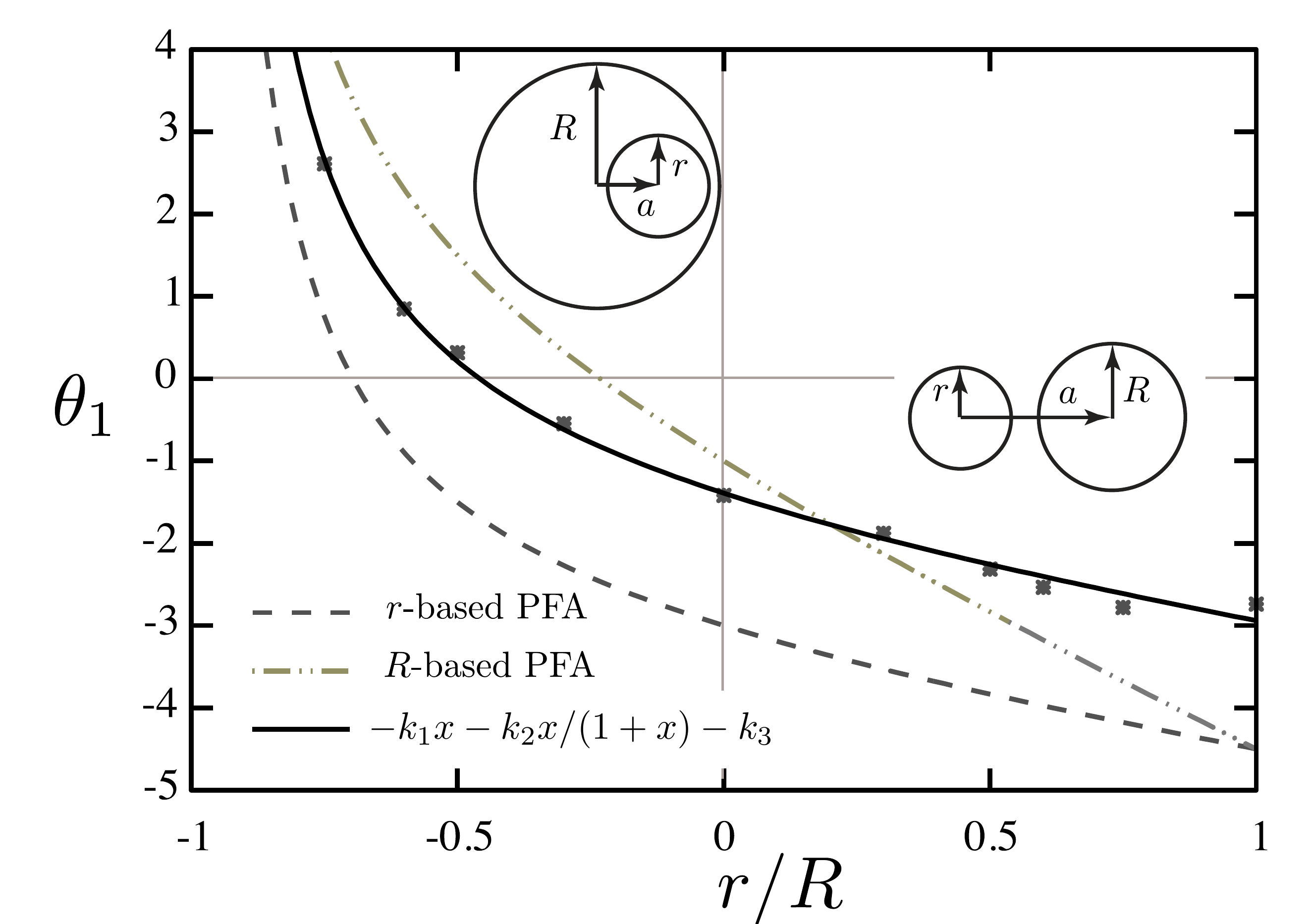}\caption{PFA correction coefficients for spheres. $r/R$ ranges from -1 (interior concentric), to zero (sphere-plane), to $+1$ (exterior, equal radii).  The data points correspond to the exact values of $\theta_1$, calculated numerically, while the solid black curve is a fit (see text). Inset: ``interior'' and ``exterior'' geometrical configurations.}
\label{pfacorrections}
\end{figure}

The above discussion suggests that corrections to the PFA extend smoothly from the interior case, $-1<r/R<0$, to the planar case, $r/R=0$, to the exterior case, $0<r/R<1$. Results for the special cases $r/R=0$ and $r/R=1$ were presented in Refs.~\cite{thorsten} and~\cite{thorstencyl}, respectively. In order to give a full description of the leading correction to the PFA, we compute the correction for several additional exterior configurations ($r/R>0$) and combine those results with the results of Refs.~\cite{thorsten} and \cite{thorstencyl}, and with our results for interior configurations in order to obtain a form for the PFA corrections over the entire possible range of $r/R$. Fig.~\ref{pfacorrections} displays our results along with the corresponding ``$r$- and $R$-based'' estimates of $\theta_1$.  
 
The numerical data in Fig.~\ref{pfacorrections} show a smooth transition from the interior to the exterior configuration. Although neither of the ``full PFA'' estimates describes the data, the $r$-based PFA has a similar functional form and divergence as $x\to -1$. Therefore, we fit the data in Fig.~\ref{pfacorrections} to a form motivated by the $r$-based PFA, $\theta_1(x) = -(k_1 x +k_2x/(1+x) +k_3)$ and find $k_1=1.05 \pm 0.14$, $k_2 = 1.08 \pm 0.08$, and $k_3 = 1.38 \pm 0.06$. This provides a simple form for the leading PFA correction for metallic spheres, one inside the other and both
outside, which is relevant for many experiments. Notice however, that the actual function $\theta_1(x)$ is not known analytically and that our fit represents a reasonable choice which may not be unique.
Our results show that the correction to the PFA has a significant dependence on the ratio of curvatures of the two surfaces. The correction is a factor of two larger for two spheres of equal radii than for the sphere-plane setup; it vanishes near $r/R=-0.5$;
and it becomes positive and large as $r/R\to -1$.  These effects should be taken into account when experimental accuracy has advanced to the point that corrections to the PFA can be measured. 

\section{Casimir-Polder limit of the interior problem}
\label{casimirpolder}

In this section, we derive the Casimir energy of a small polarizable object contained inside a metallic spherical shell. Using the matrix identity $\ln \det \mathbb{M} = \text{Tr} \ln\mathbb{M}$ we expand the integrand in Eq.~(\ref{eq:master}) in a Taylor series, ${\cal E}= \hbar c/2\pi \int d\kappa{\rm Tr}\left( \mathcal{N} + \frac{1}{2}\mathcal{N}^{2}+...\right)$, in $\mathcal{N} = \mathcal{T}_{e}^{ii}  \mathcal{V}_{e,i}  \mathcal{T}_{i}^{ee}  \mathcal{V}_{i, e}$ where each matrix multiplication by $\mathcal{N}$ describes the propagation and reflection of a virtual photon from the pair of objects~\cite{thorsten}. In the exterior case, both objects can be taken to be small compared to the separation between them, and can be approximated by their 
static dipole polarizabilities in this limit, leading to the famous Casimir-Polder force between polarizable molecules.  In the interior case studied here, the enclosing cavity must be larger than the interior object, and cannot be approximated by its lowest frequency electromagnetic response. Therefore, any asymptotic expansion of Eq.~(\ref{eq:master}) requires that the shape and material properties of the enclosing cavity be specified and its $\cT$-matrix be calculable. We obtain a useful expansion by requiring the size of the internal object to be small enough that it be adequately  described by its electric and magnetic dipole polarizability tensors. Thus the most general interior configuration to which an analytic 
expansion applies is that of a small dielectric object immersed in a dielectric medium inside a dielectric cavity for which the scattering amplitudes are known analytically in some partial wave basis.
Our results on the Casimir-Polder limit of a small internal object inside a metallic spherical shell were reported in Ref.~\cite{Zaheerletter}, and stable three dimensional configurations of objects in spheroidal drops of liquid or metallic shells were presented in Ref.~\cite{Jamal09}. 
\begin{figure}[t] 
\includegraphics[width= 10cm]{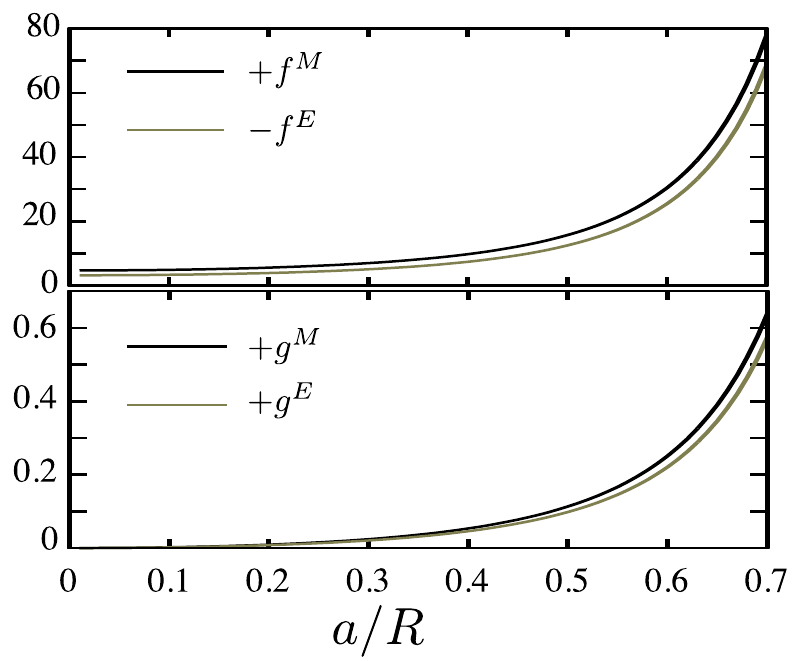}\caption{\label{dfunction}Plot of the functions $f^{M/E}(a/R)$ and $g^{M/E}(a/R)$, defined in Eqs.~(\ref{eq:trace1}) and (\ref{eq:trace2}), respectively.}
\label{hdme}
\end{figure}

Here we consider a metallic spherical cavity of radius $R$ enclosing a small object of typical linear dimension $r$, and use the $\cT$-matrix representations in Eqs.~(\ref{eq:tmatrix}). Substituting the lowest order approximation to the interior object's $\cT$-matrix, $\cT^{ee}_{i}$, but keeping all partial waves in the $\cT$-matrix of the exterior sphere, $\cT^{ii}_{e}$, we find the first term in the expansion in $r/R$, with coefficients that are non-trivial functions of $a/R$. This expansion was already quoted in the Introduction,
\begin{align}
  &\frac{3\pi R^{4}}{\hbar c} \mathcal{E}(a/R)     = \left[f^{E}(a /R)-f^E(0)\right]  {\text{Tr} \alpha^E}  + g^{E}(a/R) (2 \alpha^E_{zz}-\alpha^E_{xx}- \alpha^E_{yy})   + (E \leftrightarrow M) + \ldots \label{eq:orientation1}
\end{align}
where $a$ denotes the displacement of the internal object along the $\mathbf{z}$-axis from the center of the cavity and  ${\alpha}$ denotes the dipole polarizability tensor expressed in a Cartesian basis. The coefficient functions $f^P$ and $g^P$, plotted in Fig.~\ref{hdme}, can be
expressed in terms of modified spherical Bessel functions $i_\nu$ and $k_\nu$ as\sjrdel{,}

\begin{align}
f^E(\xi) &= \int_0^\infty dx x^3 \sum_{l=1}^{\infty} \Bigg[ \frac{\zeta_l^E(x)}{2} \left((l+1)i^2_{l-1}(x \xi) + li^2_{l+1}(x \xi)\right) -\zeta_l^M(x) \frac{ x^2\xi^2}{2(2l+1)}\left(i_{l-1}(x \xi) - i_{l+1}(x \xi)\right)^2\Bigg] \, , \label{eq:trace1}
\end{align}
\begin{align}
g^E(\xi) = \int_0^\infty dx x^3 \sum_{l=1}^{\infty} &\Bigg[\frac{\zeta_l^E(x)}{2(2l+1)}\Big(\frac{l^2-1}{2}i^2_{l-1}(x \xi) + \frac{l(l+2)}{2}i^2_{l+1}(x \xi) - 3l(l+1)i_{l-1}(x \xi)i_{l+1}(x \xi) \Big) 
\nonumber \\ &+ \zeta_l^M(x)\frac{x^2 \xi^2}{4(2l+1)} \left(i_{l-1}(x \xi) - i_{l+1}(x \xi)\right)^2\Bigg] \, ,\label{eq:trace2}
\end{align}
and $f^M$ and $g^M$ are obtained by substituting $(E \leftrightarrow M)$ in the above equations. The functions $\zeta_l^{M/E}$ are given by
\begin{align}
\zeta_l^M(x) = \frac{k_l(x)}{ i_l(x)}, \qquad
\zeta_l^E(x) = \frac{k_{l}(x) + xk'_{l}(x)}{i_{l}(x) + xi'_{l}(x)} \, .\label{eq:gamma}
\end{align}
 
An analogous result --- a series in $r/R$ with coefficients that are non-trivial functions of $a/R$ --- can be obtained for an analogous ``exterior'' configuration: a polarizable object facing a metallic sphere of radius $R$, see Ref.~\cite{Zaheerletter}. Both results differ from the classic Casimir-Polder result, which describes an object facing a conducting plane, in three ways: $f,g$ are non-trivial functions of $a/R$; the polarizable object experiences a torque; and the Casimir force between the two objects at leading order depends on the polarizable object's orientation. Specific features of the above orientation dependence are explored by specifying the internal object to be a dielectric spheroid in Ref.~\cite{Zaheerletter}. For example, it is demonstrated that the orientation dependence, like the leading PFA correction, has a smooth continuation from the interior to the exterior configurations; a `cigar-shaped' spheroid prefers to align itself perpendicular to its displacement vector from the center of the shell; and a `pancake-shaped' spheroid prefers to align its two large axes perpendicular to its displacement vector. 
\begin{figure}[t]
\includegraphics[scale=0.5]{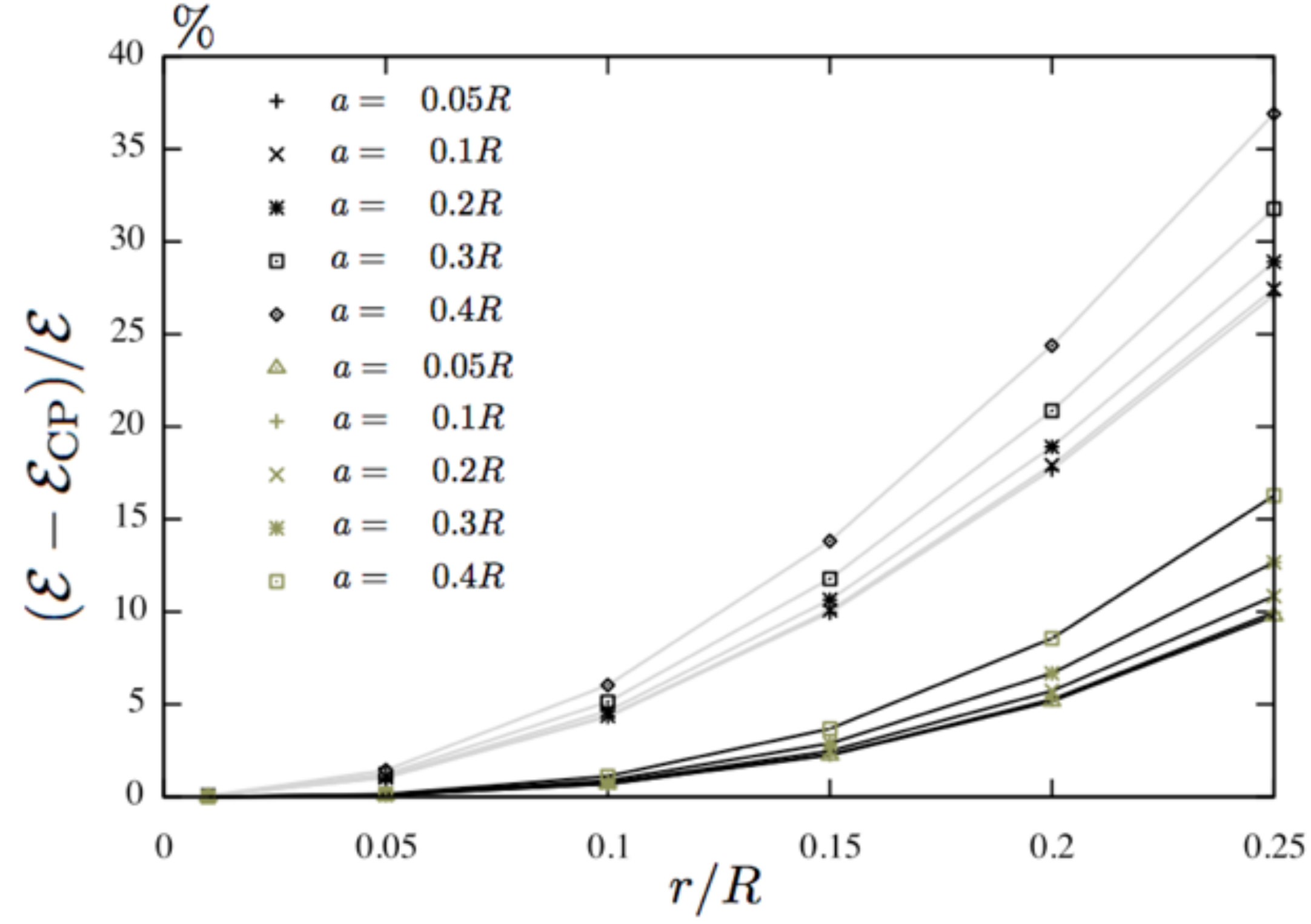} \caption{Comparison of the interior Casimir-Polder result with the exact Casimir energy predicted by Eq.~(\ref{eq:master}) for conducting spheres. Plotted along the $y$-axis is the fractional error $(\cE - \cE_{CP})/\cE$ as a percentage, where $\cE_{CP}$ is calculated from Eq.~(\ref{eq:casimirpolder}). For each value of the separation $a/R$ (denoted by point markers listed above), we have plotted the fractional difference between $\cE$ and $\cE_{CP}$ at the leading order $\cO(r^3/R^3)$ (gray) and next-to-leading order $\cO(r^5/R^5)$ (dark black) as a function of the internal sphere radius $r/R$. 
\label{vectorerrors}}
\end{figure}

For a spherically symmetric internal object, the first non-trivial correction, which is ${\cal O}(r^{5}/R^{5})$, to Eq.~(\ref{eq:orientation1}) can be easily evaluated. This is possible because the $\cT$-matrix for the spherically symmetric object is diagonal in a spherical basis and independent of the azimuthal index, $m$, 
${\cal T}_{lml'm'}^{\lambda\sigma}=\delta_{ll'}\delta_{mm'}\delta^{\lambda\sigma}{\cal T}_{l}^{\sigma}$.   For each polarization and $l$, the leading term in the $\cT$-matrix is proportional to a multipole polarizability, $\alpha^{\sigma}_{l}\sim r^{2l+1}$, which characterizes the low frequency electromagnetic response of the internal object, and which can be computed for simple geometries or measured for any compact conductor~\cite{thorsten},
\begin{align}
\mathcal{T}_{l}^{\sigma} & = \kappa^{2l} \Bigg[ \frac{(-1)^{l-1} (l+1) \alpha_l^\sigma}{l (2l+1)!! (2l-1)!!} \kappa  + \cO(\kappa^3) \Bigg] \label{eq:smalltmatrix}
\end{align} 
for $\sigma=E$ or $M$. Substituting Eqs.~(\ref{eq:tmatrix}) and~(\ref{eq:vmatrix}) in $\text{Tr }\mathbb{N}$ we find the Casimir energy up to $\mathcal{O}(r^5/R^5)$ to be, 
\begin{align}
\frac{2 \pi R}{\hbar c}\mathcal{E}_{\rm CP} &= h_1^M(a/R) \frac{\alpha_1^M}{R^3} + h_2^M(a/R)\frac{\alpha_2^M}{R^5} + (M\leftrightarrow E)+ \cO(r^6/R^6) \label{eq:casimirpolder}
\end{align}
where $h^{M/E}_1$ are proportional to $f^{M/E}(a/R)$ defined in Eq.~(\ref{eq:trace1}) up to a numerical factor. The exact functional forms of the coefficient functions $h_1$ and $h_2$ are given in the Appendix. Each can be expanded as a power series in $a^2/R^2$ which converges for $a^2/R^2 <1$. The Casimir force can be calculated by differentiating  $h_1, h_2$ with respect to $a$. 

To examine the usefulness of the expansion in Eq.~(\ref{eq:casimirpolder}), we compare its predictions with the exact numerical result for $\cE$ following from Eq.~(\ref{eq:master}) for the case of an internal metallic sphere of radius $r$, for which $\alpha^M_l = -l r^{2l+1}/(l+1)$ and $\alpha^E_l = r^{2l+1}$. Fig.~\ref{vectorerrors} plots the fractional errors $\Delta \cE = (\cE - \cE_{CP})/\cE$ vs. $r/R$ for various separations $a/R$. The 1st-order data (in black markers) include contributions from $\cO(R^{-3})$ terms only, while the 2nd-order data (in gray markers) include coefficients from $h_2(a/R)$ at $\cO(R^{-5})$ in Eq.~(\ref{eq:casimirpolder}). Many trends are visible in this graph:  For example the interior Casimir-Polder result through second order is accurate to more than $99\%$ of the exact answer for all $r/R \leq 0.1$ for $0<a/R< 0.4$. Another feature worth noting is that for a given value of $r/R$, $\lim_{a/R \to 0} \Delta \cE$ is not zero. Both the exact Casimir energy, ${\cal E}$, and the interior Casimir-Polder approximation, ${\cal E}_{\rm CP}$, vanish like $a^{2}$ as $a\to 0$. (Remember the value of each at $a=0$ has been subtracted.)  Notice however, that the Casimir Polder approximation is an expansion in $r/R$, not $a/R$, and therefore each term in the expansion, Eq.~(\ref{eq:casimirpolder}), contributes, albeit with smaller magnitude, at $a=0$. Fig.~\ref{vectorerrors} shows that the interior Casimir-Polder expansion becomes exact for an arbitrarily small polarizable object (an atom or a molecule) inside a conducting spherical shell.

\section{Conclusions}
We have studied the electromagnetic Casimir problem for a compact object contained inside a closed cavity of another compact object. Using the scattering formalism, we express the Casimir energy between the two objects in terms of their $\mathcal{T}$-matrices, and translation matrices that relate the coordinate systems appropriate to each object. Then we specialize to the case when both objects are conducting spheres, and illustrate our methods and results by evaluating the Casimir energy for the case $r/R = 0.5$. The Casimir force for this sphere configuration is calculated by numerically differentiating the energy with respect to the spheres' separation. 

We have also calculated the analog of the Casimir-Polder expansion for an object contained inside a metallic spherical shell. The Casimir energy can be expanded as an asymptotic series in $r/R$ (the leading term being proportional to $r^3/R^3$), where $R$ is the radius of the spherical cavity, and $r$ is a length characterizing the size of the internal object. There are certain novel features of this result: the coefficients are non-trivial functions of $a/R$ (where $a$ denotes the separation of the center of the internal object from the center of the shell) that is, they are represented by infinite sums of modified spherical bessel functions; the Casimir force at leading order depends on the orientation of the internal object; and the internal object experiences a torque. Additionally, we have calculated the coefficient functions for the leading two terms for the case when the inner object has spherical symmetry. A comparison of the `interior Casimir-Polder' expansion (up to $\cO(R^{-5})$) with the exact energy (calculated numerically) for various sphere configurations shows it to be accurate to more than $99\%$ of the exact answer for $a/R \leq 0.4$ through $r/R \leq 0.1$ where $a$ is the displacement of their centers.

The methods demonstrated in this paper can be applied very easily to calculate the Casimir force between dielectric spheres at all separations, although they become computationally intensive when the spheres are nearly touching.  In this limit, we make contact with the Proximity Force Approximation (PFA). A careful examination of the approach to the PFA allows us to calculate leading order corrections to the PFA limit.  We then combine our studies of various interior sphere configurations at close separations with previous work on metallic spheres exterior to each other and a conducting sphere facing a mirror to analyze the leading order PFA corrections for all sphere-sphere configurations.   

This work was supported by the NSF through grant DMR-08-03315 (SJR), DFG through grant EM70/3 (TE), the U. S. Department of Energy (D.O.E.) under cooperative research agreement \#DF-FC02-94ER40818 (RLJ), and MIT's undergraduate research opportunities
program (UROP) (SZ). We thank Noah Graham and Mehran Kardar for many useful conversations and suggestions regarding this work.

\appendix
\section{Interior Casimir-Polder coefficient functions}
We give the coefficient functions $h_1^M$ and $h_2^M$ that appear in Eq.~(\ref{eq:casimirpolder}) in terms of $\zeta_l^M$ and $\zeta_l^E$ defined in Eq.~(\ref{eq:gamma}) and modified spherical bessel function $i_l$. $h_1^E$ and $h_2^E$ can be obtained by substituting $M\leftrightarrow E$ in the following equations.
\begin{align}
h_1^M(\xi) &=  \int_0^{\infty}  dx  x^3\left( \sum_{l=1}^{\infty} \left\{\zeta_l^M(x) \left((l+1)i^2_{l-1}(x\xi) + li^2_{l+1}(x\xi)\right) - \zeta_l^E(x)\frac{x^2\xi^2}{2l+1}  \left(i_{l-1}(x\xi) - i_{l+1}(x\xi) \right)^2 \right\} -  2\zeta_1^M(x) \right)\end{align}
\begin{align}
h_2^M(\xi) &= \int_0^{\infty} dx x^5 \Bigg( \sum_{l=1}^{\infty}\bigg[  \zeta_l^M(x) \frac{ (l-1) (l+1) (2 l+3) i^2_{l-2}(x\xi)+l (l+2) (2 l-1)i^2_{l+2}(x\xi)+
   (3l+3/2)i^2_l(x\xi)}{6(4 l (l+1)-3)}  \nonumber \\ 
& - \frac{x^2\xi^2}{3(2l+1)} \zeta_l^E(x)\bigg\{\frac{1}{4} \left(\frac{1-l}{2l-1}i_{l-2}(x\xi) -\frac{2l-1}{4l(l+1)-3}i_l(x\xi) + \frac{l+2}{2l+3}i_{l+2}(x\xi)\right)^2  \nonumber \\ 
&+ (l-1)(l+2)\left( \frac{1}{2(2l-1)}i_{l-2}(x\xi) - \frac{2l+1}{4l(l+1)-3} i_l(x\xi) +\frac{1}{2(2l+3)}i_{l+2}(x\xi)\right)^2\bigg\}\bigg] - \frac{1}{6} \zeta_2^M(x)\Bigg)
\end{align}


\begin{thebibliography}{99}
\bibitem{casimir}
H. B. G. Casimir, Indag. Math. {\bf 10}, 261 (1948) [Kon. Ned. Akad. Wetensch. Proc. {\bf 51}, 793 (1948)]. 
\bibitem{Jamaluniversal}
S. J. Rahi, T. Emig, N. Graham, R. L. Jaffe, and M. Kardar, Phys. Rev. D {\bf 80}, 085021 (2009).
\bibitem{Jamal09}
S. J. Rahi and S. Zaheer, Phys. Rev. Lett. {\bf 104}, 070405 (2010).
\bibitem{thorsten}
T. Emig, N. Graham, R.~L. Jaffe, and M. Kardar,
Phys. Rev. Lett. {\bf 99}, 170403 (2007).
\bibitem{thorstencyl}
T Emig J. Stat. Mech. (2008) P04007. 
\bibitem{beyondpfa} See for example, H.~Gies, K.~Langfeld and L.~Moyaerts,
  J. High Energy Phys. 06 {\bf 2003}, 018;  R.~L.~Jaffe and A.~Scardicchio,
  Phys.\ Rev.\ Lett.\  {\bf 92}, 070402 (2004);
 H. Gies and K. Klingm{\"u}ller,
  Phys. Rev. Lett. {\bf 96}, 220401 (2006); D. E. Krause, R. S. Decca,
  D. Lopez, and E. Fischbach, Phys. Rev. Lett. {\bf 98}, 050403
  (2007); M. Bordag and V. Nikolaev, J. Phys. A-Math. Theor., {\bf 41}, 164002 (2008); S. Reynaud,
  P. A. Maia Neto and A. Lambrecht,  J. Phys. A-Math. Theor., {\bf 41}, 164004 (2008). 
\bibitem{bordagpfa}
M. Bordag and V. Nikolaev, arXiv:0911.0146v1 [hep-th].
\bibitem{Zaheerletter}
S. Zaheer, S. J. Rahi, T. Emig, and R. L. Jaffe, Phys. Rev. A {\bf 81}, 030502(R) (2010).
\bibitem{marachevsky2001}
Valery N. Marachevsky, Mod. Phys. Lett. A Vol. 16 No. 15 (2001) 1007-1016;  Valery N. Marachevsky, Phys. Scr. 64 205 (2001). 
\bibitem{plates}
Valery N. Marachevsky, Phys. Rev. D {\bf 75}, 085019 (2007) 

\bibitem{brevik02}
I. Brevik, J. B. Aarseth, and J. S. Hoye, Phys. Rev. E {\bf 66}, 026119 (2002); J. S. Hoye, I. Brevik, and J. B. Aarseth, Phys. Rev. E {\bf 63}, 051101 (2001).
\bibitem{cylinders}
D. A. R. Dalvit, F. C. Lombardo, F. D. Mazzitelli, and R. Onofrio,
Phys. Rev. A {\bf 74}, 020101(R) (2006).
\bibitem{KK} O. Kenneth and I. Klich, Phys. Rev. B {\bf 78}, 014103 (2008).
\bibitem{Maia_Neto} P.~A.~Maia~Neto, A.~Lambrecht, and S.~Reynaud, Phys. Rev.~A {\bf 78}, 012115 (2008).
\bibitem{kardar}
H. Li and M. Kardar, Phys. Rev. Lett. {\bf 67}, 3275 (1991); Phys. Rev. A {\bf 46}, 6490 (1992).
\bibitem{scalar}
T. Emig, N. Graham, R.~L. Jaffe, and M. Kardar, Phys. Rev. D {\bf 77}, 025005 (2008). 
\bibitem{orientation}
T. Emig, N. Graham, R. L. Jaffe, and M. Kardar, Phys. Rev. A {\bf 79}, 054901 (2009).
\bibitem{casimirpolder} H.~B.~G. Casimir and D. Polder,  Phys. Rev. {\bf 73}, 360 (1948).
\bibitem{PFA}
B. V. Derjagin, Kolloid Z. {\bf 69} 155 (1934), B. V. Derjagin, I. I. Abriksova, and E. M. Lifshitz, Sov. Phys. JETP {\bf 3}, 819 (1957); For a modern discussion of the Proximity Force Theorem, see J. Blocki and W. J. Swiatecki, Ann. Phys. {\bf 132}, 53 (1981).
\bibitem{Schaden:1998zz}
  M.~Schaden and L.~Spruch,
  Phys.\ Rev.\  A {\bf 58}, 935 (1998).
\bibitem{thesis}
S. Zaheer, Undergraduate thesis, Massachusetts Institute of Technology, 2009.
\bibitem{wittmann}
R. C. Wittmann, IEEE Transactions on Antennas and Propagation, {\bf 36}, 1078 (1988) .
\end{thebibliography}
\end{document}